\documentstyle[prl,aps,multicol,epsf]{revtex} 
\begin{document}    

\draft   

\twocolumn[ 
\hsize\textwidth\columnwidth\hsize\csname@twocolumnfalse\endcsname 

\title{3D Lowest Landau Level Theory Applied to YBCO Magnetization and Specific Heat Data: 
Implications for the Critical Behavior in the $H$-$T$ Plane}

\author{Stephen W.~Pierson$^1$\cite{email}, Oriol T.~Valls$^2$, Zlatko Te\v sanovi\' c$^3$, and
Michael A.~Lindemann$^1$}    
\address{$^1$Department of Physics, Worcester Polytechnic
Institute, Worcester,    MA 01609-2280}    
\address{$^2$School of Physics and Astronomy and
Minnesota Supercomputer Institute \\  University of Minnesota, Minneapolis, MN 55455-0149}  
\address{$^3$Department of Physics and Astronomy, Johns Hopkins University, Baltimore, MD 21218}
\date{\today}    
\maketitle    
\begin{abstract} 
We study the applicability of magnetization and
specific heat equations derived from a lowest-Landau-level (LLL) calculation, to the
high-temperature superconducting (HTSC) materials of the YBa$_2$Cu$_3$O$_{7-\delta}$ (YBCO)
family. We find that significant information about these materials can be obtained from this
analysis, even though the three-dimensional LLL functions are not quite as successful in
describing them as the  corresponding two-dimensional  functions are in describing data for the
more anisotropic HTSC Bi- and Tl-based materials.  The results discussed include scaling fits,
an alternative explanation for data claimed as evidence for a second order flux lattice melting
transition, and reasons why 3DXY scaling may have less significance than previously believed. We
also demonstrate how 3DXY scaling does not describe the specific heat data of YBCO samples in the
critical region. Throughout the paper, the importance of checking the actual scaling functions,
not merely scaling behavior, is stressed.

\end{abstract}    
] 

\narrowtext    
\section{Introduction}    
\label{sec:intro}  
The critical behavior of the high-temperature superconductors in finite magnetic fields applied
perpendicular to the copper oxide planes has been described by both lowest Landau level (LLL)
theory and three-dimensional (3D) XY theory with varying degrees of success. It is widely
expected that 3DXY behavior should hold at low fields and that LLL should be valid at higher
fields. There is however little consensus about what the value of the crossover field should be
and how well either of these theories describe resistivity data, magnetization data, or specific
heat data\cite{leeds94a,pierson95com,roulin95,roulin96}. One group\cite{moloni96} has 
claimed that LLL should not be valid for fields less than ten Tesla (T) in deoxygenated
YBa$_2$Cu$_3$O$_{7-\delta}$ (YBCO) thin films based on conductivity measurements while three of
the present authors\cite{pierson96,zhou93} have found it to be valid down to approximately two
Tesla based on  an analysis of specific heat data from YBCO and LuBa$_2$Cu$_3$O$_{7-\delta}$
(LBCO) single crystals.

On the theoretical front, considerable work has been done--namely,
the derivation of analytic expressions for the
magnetization and specific heat LLL
scaling functions for two dimensional (2D), three-dimensional (3D), 
and layered systems. This was achieved \cite{zlatko92,zlatko94} by using
the LLL approximation in the Ginzburg-Landau (GL) formalism. The
two-dimensional portion of this work has had 
striking success in describing the {\it magnetization} of the
highly anisotropic 
high-temperature superconducting materials Bi$_2$Sr$_2$CaCu$_2$O$_8$ (BSCCO-2212) and the
Tl-based compounds for magnetic fields applied perpendicular to the copper oxide planes. For
example, the two-dimensional (2D) functions\cite{zlatko92,zlatko94} have a field independent
value at a particular temperature $T^*$\cite{bulaevskii92} which is also a crossing point for
the magnetization curves. Such behavior has been observed in BSCCO-2212 by many
authors\cite{kes91} and furthermore, Wahl and coworkers\cite{wahl95} have not only observed such
crossover in their magnetization data from Tl-based single crystals but they have also fit the
2D functions of Ref.~\onlinecite{zlatko92,zlatko94} to their data finding good agreement. 

Little work has been done to fit the theoretical functions to {\it specific heat} data on
the highly anisotropic HTSC materials. Kobayashi {\it et al.}\cite{koba94} are among the few to
publish specific heat data\cite{tonyc96} for various fields near the critical temperature on such 
compounds.
They scaled their specific heat data from a c-axis aligned (Bi,Pb)$_2$Sr$_2$Ca$_2$Cu$_3$O$_x$
bulk sample and compared it to the 2D scaling function of Te\v {s}anovi\' c and
coworkers\cite{zlatko92,zlatko94} finding reasonable agreement. They also found a crossing point
in their magnetization data.

Even less has been done to compare 
the theoretical expressions and
scaling functions to experimental data for the more isotropic YBCO materials. 
In Ref.~\onlinecite{pierson96}, an approximation to the 3D
LLL specific heat function was compared 
to scaled specific heat data from various YBCO samples (including a YBCO single 
crystal from Ref.~\onlinecite{leeds94a}) and a LBCO sample,
with satisfactory agreement. Further, 
in the work of Ref.~\onlinecite{roulin96}, although a quantitative comparison was not made, one
can find qualitative agreement between the scaled temperature derivatives of the specific heat
and the second derivative of the 3D magnetization function. (See Eq.~(5) and Figs.~6 and 8 of
that reference.) Lastly, we are not aware of any work comparing the 3D theoretical magnetization
function to magnetization data of YBCO-class materials. This can be attributed in part to the
complexity \cite{zldetail} of the 3D specific heat and magnetization functions of
Ref.~\onlinecite{zlatko94}.

In this paper, we examine the 3D specific heat and magnetization functions of 
Ref.~\onlinecite{zlatko94} comparing them to data from YBCO samples. Not only is such a 
comparison lacking and certainly needed in order to learn more about the
validity of the theory but we will see that
it yields valuable insights into other questions about the behavior of the
YBCO materials, besides the nature of their fluctuation behavior. For example, we will 
demonstrate how data presented 
as evidence for a second order flux lattice melting transition can be explained within LLL
theory without invoking flux lattice melting arguments. Furthermore, it will also be seen that
the 3DXY scaling is so general as to describe ``theoretical data'' derived using the LLL theory.
This exemplifies the importance of knowing the actual scaling function and leads one to question
the significance of 3DXY scaling. The failure of 3DXY scaling to describe specific heat data
from YBCO samples will also be shown. 

Our focus here will not be so much on the validity of LLL scaling for the HTSC materials 
as on the applicability of the specific expressions and scaling functions from one 
particular calculation based on a nonperturbative approach to the GL-LLL 
theory\cite{zlatko92,zlatko94}. There are two separate issues here. First, there is the
question of describing the HTSC's in a GL-LLL formalism, which has already been answered, in our
opinion, through the success of LLL scaling. Second, while the expressions we use are of compact
form and should be useful for analysis of experiments and phenomenology, they are not
mathematically identical to the exact solution of the GL-LLL theory. Therefore, there
is a need to address any possible disagreement between theory and experiment arising from the
additional approximations involved in the expressions for the scaling functions of
Refs.~\onlinecite{zlatko92,zlatko94} relative to the exact answer within the GL-LLL theory.
Although this issue is essentially resolved for very anisotropic, ``almost" 2D HTSC systems
where the 2D LLL scaling functions of Refs.~\onlinecite{zlatko92,zlatko94} are known to be
very accurate,\cite{footi} it has not been investigated for the relatively isotropic materials
of the YBCO class. 

In this work, we take advantage of the availability of numerical work 
on the (quasi) 3D GL-LLL model\cite{sasik95}. The existence of this numerical work
for the magnetization will allow us, as we shall see, to determine some
of the fitting parameters in a way that it is not constrained with
experimental uncertainties involving, for example, the subtraction
of ``background'' terms. We can say that we use these numerical 
results to ``calibrate'' certain parameters in the scaling functions.
This is very convenient, since the increased complexity of the 3D
functions, as opposed to the 2D case, would otherwise make our task
much more intricate and the conclusions weaker.

The paper is organized as follows: The theoretical functions calculated from the 
nonperturbative approach\cite{zlatko92,zlatko94} to the GL-LLL theory 
will be set forth and discussed in Section \ref{sec:th}. In Section \ref{sec:mag}, the 
``calibration'' fits of the numerical 3D magnetization data to the theoretical result, 
[Eq.~(\ref{M})] are performed, and then fits to actual magnetization data from YBCO samples 
are done. Then, in \ref{sec:der}, we give a simple explanation of the peculiar behavior of 
the field dependence of the partial derivative $\partial M(H,T) / \partial T$ 
found\cite{jean96b} in YBCO 
and BSCCO. We show that this behavior is simply explained in terms of the LLL scaling functions.
Fits of the theoretical 3D specific heat function [Eq.~(\ref{SH})] to specific heat data from
the same materials as in \ref{sec:mag} are reported in Section \ref{sec:sh} along with an
alternative explanation to data claimed as evidence for a second order flux lattice melting
transition. Finally, implications of this work for 3DXY scaling and the importance of the
scaling functions is demonstrated in Section \ref{sec:3dxy} followed by a discussion and summary
in Section \ref{sec:summ}.

\section{Theoretical Functions and Data Fits} 
\label{sec:body}
\subsection{Theoretical GL-LLL Functions}    
\label{sec:th}
As mentioned above, the 2D functions of Refs.~\onlinecite{zlatko92,zlatko94}
have had considerable success in describing magnetization data from the
highly anisotropic HTSC materials. Here, we will focus on the 3D specific
heat and magnetization functions since the 2D functions are already 
examined and the quasi-2D functions are less tractable. The magnetization as a function of
applied magnetic field $H$ and temperature $T$ is written as Eq.~(26) of
Ref.~\onlinecite{zlatko94} 
\begin{eqnarray} 
&&{4\pi M(H,T) H_{c2}'\over (TH)^{2/3}}  \Biggl
({4\sqrt{2\pi T_{c0}}\xi \kappa \phi_0\over k_B H_{c2}'}\Biggr )^{2/3} =\nonumber \\ 
&&\Biggl (g+\sqrt{g^2 +{tan^{-1} Q\over \pi U^2}}\Biggr )^{1/3} \Bigl [GU^2-U\sqrt{G^2U^2+2} 
\Bigr ].
\label{M} \end{eqnarray} 
The specific heat function can be rewritten (including important
subleading terms) by  considering\cite{q3d} the pure 3D limit of the quasi 3D result (Eq.~(30)
of Ref.~\onlinecite{zlatko94}): 
\begin{eqnarray} {C(H,T)\over C_{MF}(T)}=&&{1 \over 2}\Biggl (1-
{GU\over \sqrt{G^2U^2+2}}\Biggr )\nonumber \\ 
&& \Biggl [U^2 {dG\over dg} + \Biggl
(\sqrt{G^2U^2+2}-GU \Biggr )\Bigl | {dU\over dG}\Bigr |\Biggl ], \label{SH} \end{eqnarray} 
where
\begin{equation} U(G)=0.818-0.110\times {\rm tanh} \bigl ({G+K\over M}), \label{U}
\end{equation} 
\begin{equation} G= g+ I \Biggl (g+\sqrt{g^2 +{{\rm tan}^{-1} Q\over \pi
U^2}}\Biggl ), \label{G} \end{equation} 
$I=Q-{\rm tan}^{-1}Q /[2 {\rm tan}^{-1} Q]$, 
and it can be shown that 
\begin{equation} {dG\over dg}={Ig +(1+I) \sqrt{g^2 +{{\rm tan}^{-1} Q/[ \pi
U^2]}}\over \sqrt{g^2 +{{\rm tan}^{-1} Q/[ \pi U^2]}}+{I {\rm tan}^{-1}Q/[ \pi U^3]}{dU\over
dG}}. \end{equation} 
Recall that $g$ is related to the temperature through 
\begin{equation}
g\Biggl (g+ \sqrt{g^2 +{{\rm tan}^{-1} Q\over \pi U^2}}\Biggr )^{1/3}=Bt, \end{equation} 
where $B=(H_{c2}'^2 \xi  \phi_0 \sqrt{T_{c0}}/[8\pi\sqrt{2}\kappa^2 k_B] )^{2/3}$ and 
\begin{equation} t\equiv \frac{T-T_c(H)}{(HT)^{2/3}}. \label{tsc} \end{equation} 
In the above equations,\cite{eqdetail} $H_{c2}'$ is the first derivative of the critical field 
$H_{c2}(T)$ with respect to temperature (and is assumed to be a constant), 
$\kappa=\lambda_{ab}/\xi_{ab}$ (where $\lambda_{ab}$ and $\xi_{ab}$ are the penetration depth 
and coherence length respectively in the $ab$ plane), $\xi$ is $\xi_c$ to within a 
multiplicative constant ($\xi_c$ is the coherence length along the $c$-axis), $T_{c}(H)$ is 
the finite field mean-field transition temperature, $T_{c0}=T_c(0)$, $C_{MF}(T)$ is 
the mean-field specific heat, $\phi_0$ the superconducting flux quantum, $k_B$ the Boltzman 
constant, and $Q$, $K$, and $M$ are adjustable parameters whose values are roughly $\pi$,
$\sqrt{2}$ and $2\sqrt{2}$ respectively. The factor of $4\pi$ in Eq.~(\ref{M}) is needed to 
convert $M(H,T)$ which is in units of emu/cm$^3$ to Gauss (G). 

The fits to Eqs.~(\ref{M}) and (\ref{SH}) are nontrivial since $U(G)$
[Eq.~(\ref{U})] and $G(g)$ [Eq.~(\ref{G})] are functions of one another and
the temperature $T$ is related to $g$ through a transcendental equation. In place of a
commercial fitting package, we have written a code which uses IMSL routines to fit the functions
[Eqs.~(\ref{M}) and (\ref{SH})] (which must be calculated self-consistently) to the data.

The 3D scaling functions [Eqs.~(\ref{M}) and (\ref{SH})] are more complex
than those of the 2D functions because there is
an extra length scale which describes the bending 
of the vortex lines along the field direction (here taken to be the $c$-axis).
For this reason, as we noted above, we will take advantage of the numerical
results\cite{sasik95} on the GL-LLL theory for the magnetization, which is the only quantity
presently available from numerical work, to ``calibrate" certain parameters in our 3D scaling
functions.

\subsection{Magnetization}    
\label{sec:mag}
As explained above, we will begin by narrowing down the number of parameters
available to perform fits to actual experimental data by first considering 
fits to results from a numerical calculation which simulates YBCO. We will then use our 
results from this fitting as a means of ``calibrating'' Eqs.~(\ref{M}) and (\ref{SH}). 

\begin{figure}[htb]
\centerline{\epsfysize=2.4truein \epsffile{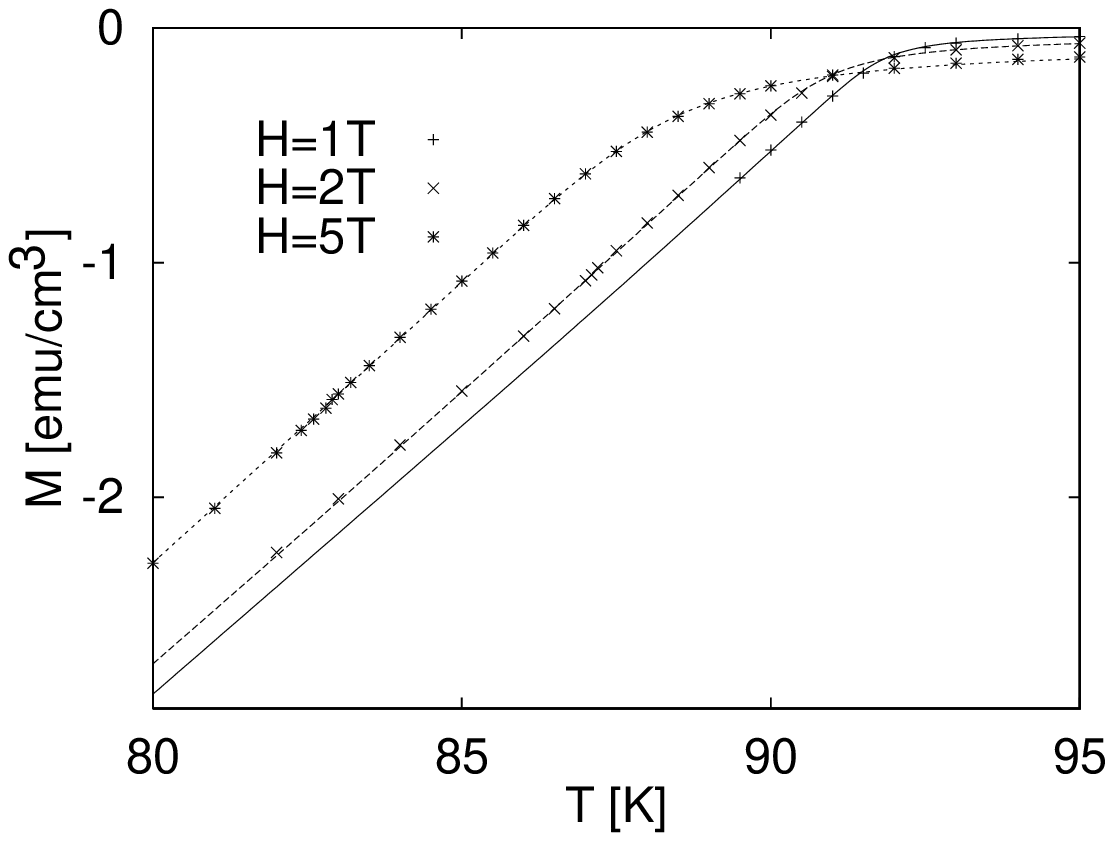}}
\centerline{\epsfysize=2.4truein \epsffile{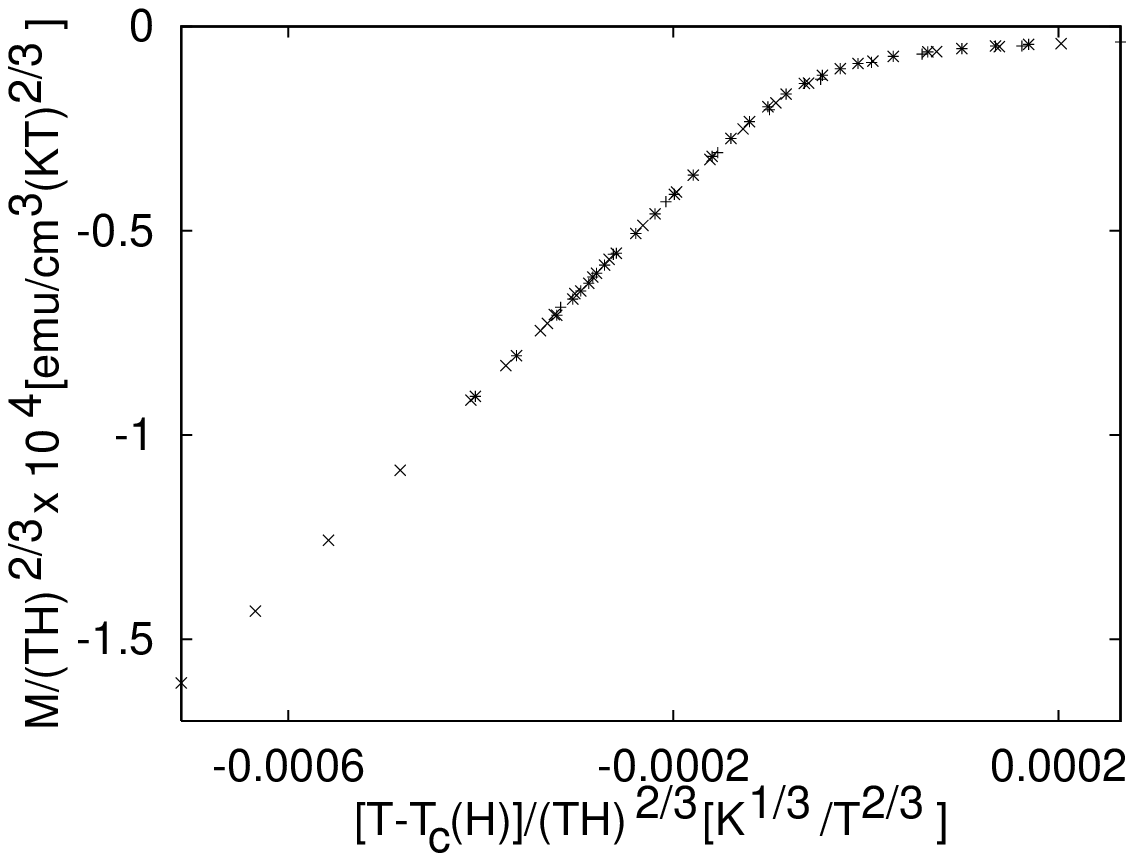}}
\caption[]{\em{
(a) A fit of Eq.~(\protect\ref{M}) to the ``numerical'' magnetization data of
Ref.~\protect\onlinecite{sasik95}. (b) The same data\protect\cite{sasik95}
scaled according to 3D LLL theory.}}
\label{sasikfig}  
\end{figure}

The numerical calculation to which we refer above was done by \v S\'a\v sik and 
Stroud\cite{sasik95} using a GL-LLL formulation for a layered system (with
parameters similar to those of YBCO $H_{c2}'=1.8 T$, $\kappa=52$, $T_{c0}=93$, and an 
anisotropy factor $\gamma=\xi_{ab}/\xi_c=5$) in order to
study flux lattice melting. Their results are reproduced in Fig.~\ref{sasikfig}a. 
We have done a four-parameter fit of Eq.~(\ref{M}) to this numerical ``data''
\cite{sasik95} and plot our results as lines (solid for 2T, dashed for 3T, and dotted for 5T)
with the data in Fig.~\ref{sasikfig}a. As one can see, the agreement is excellent. The fitting
parameters are $Q$, $K$, $M$, and the constant $A$ relating $\xi$ to $\xi_c$ and we find
$Q=10.25$, $K=-5.95$, $M=7.38$, and $\xi=0.2918\xi_{c}$ (i.e., $A=0.2918$) since $\xi_c=2.82\AA$
here. To verify the consistency of the numerical data with LLL theory, we scaled the data
according to the 3D LLL form $M(H,T)/(HT)^{2/3}=f([T-T_c(H)]/ (TH)^{2/3})$ and found the data 
to collapse flawlessly as shown in Fig.~\ref{sasikfig}b. When doing such scaling, one can
typically use $T_{c}(H)$ as an adjustable parameter (and, to a lesser extent, the background
parameters), but since $H_{c2}'$ and $T_{c0}$ are known in this case, there are no adjustable
parameters, which makes the scaling of this data very convincing. 

\begin{figure}[htb]
\centerline{\epsfysize=2.4truein \epsffile{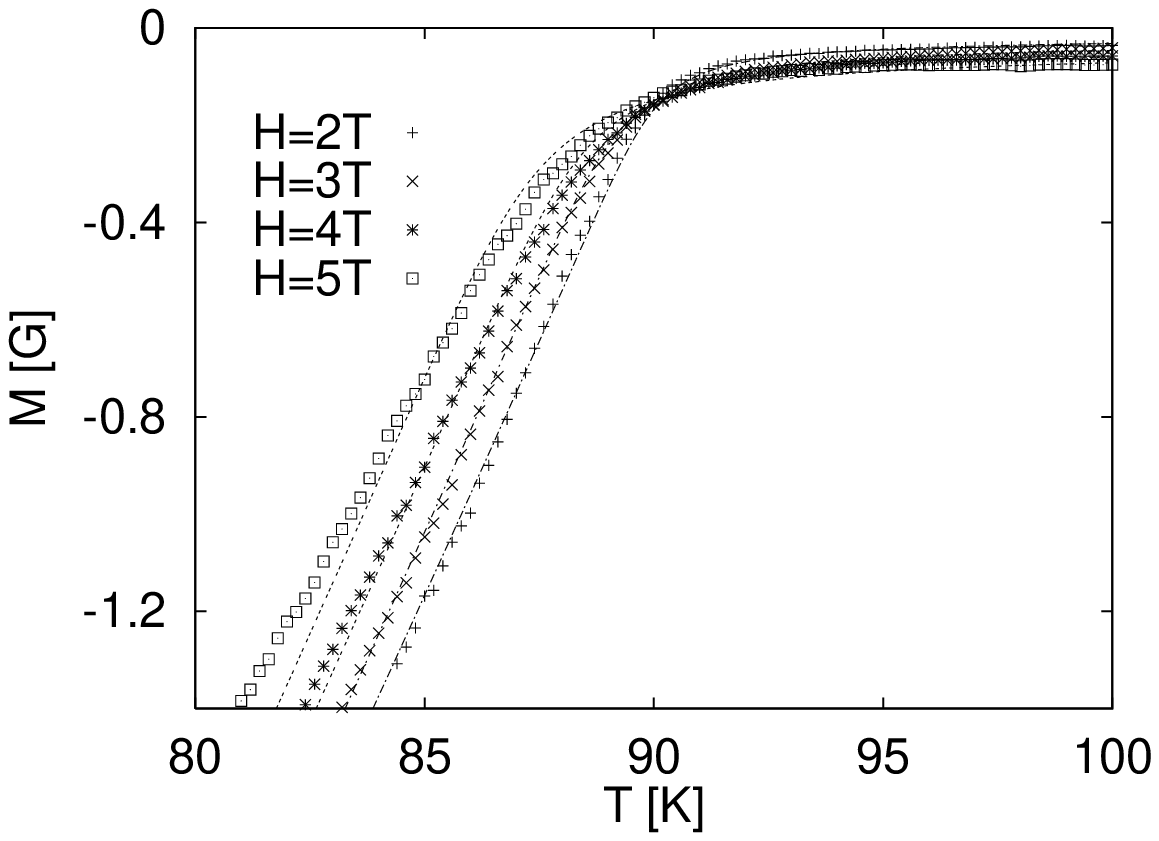}}
\centerline{\epsfysize=2.4truein \epsffile{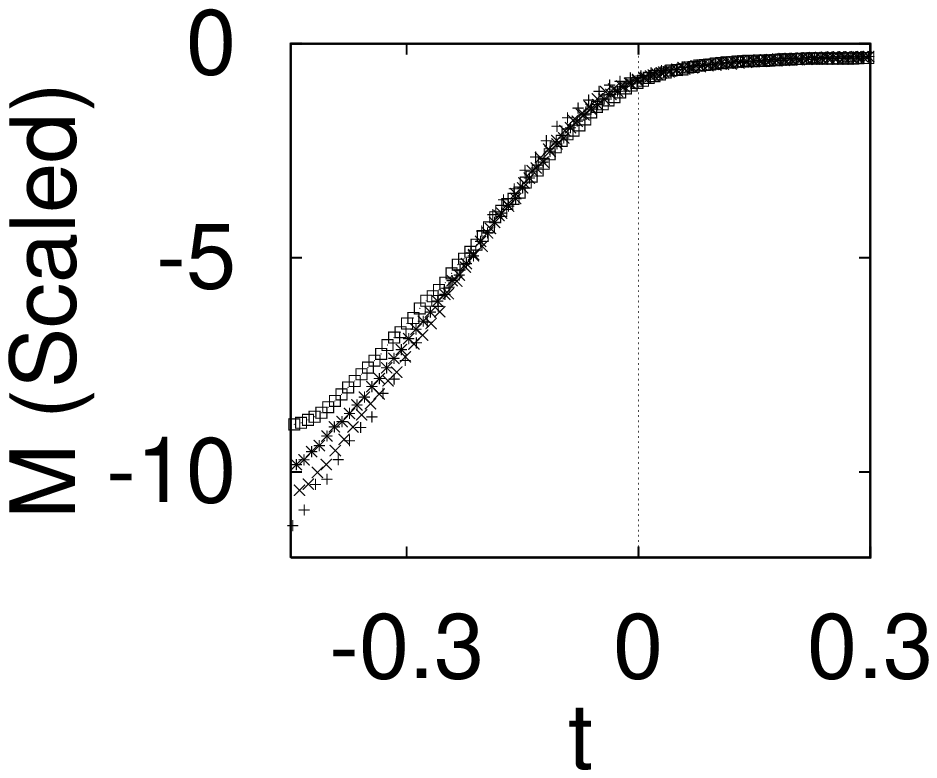}}
\caption[]{\em{
The magnetization data from a YBCO 
twinned single crystal published in Ref.~\protect\onlinecite{jean96} along with the fits
to Eq.~(\protect\ref{M}).  [INSET: Our LLL scaling of the magnetization data of 
Ref.~\protect\onlinecite{jean96} shown over a larger range than that of the authors of 
Ref.~\protect\onlinecite{jean96} which is in the Inset to Fig.~5 of that reference. The $
y$-axis 
is $4\pi M/(TH)^{2/3}\times 10^4 [G/(OeK)^{2/3}$ and the $x$-axis is  $t\times10^3 
[K^{1/3}/Oe^{2/3}]$ where $t$ is defined in Eq.~\protect\ref{tsc}.]}} 
\label{jeanfig} 
\end{figure}

Using these values of $Q$, $K$, $M$, and $A$ determined from the
fit to the numerical ``data'' we then performed
a fit to the two and three Tesla magnetization data of Jeandupeux {\it et 
al}\cite{jean96}. There are nine fitting parameters, namely
$H_{c2}'$, $\kappa$, $\xi_c$, $T_c(2T)$, $T_c(3T)$, $B_0(2T)$, $B_1(2T)$, $B_0(3T)$, and 
$B_1(3T)$ where $B_0(H)$ and $B_1(H)$ are the field-dependent constants used to adjust the
subtracted background: $M_B=(B_0+B_1/T)H$\cite{back}. Three-parameter fits were then used to
find $T_c(H)$, $B_0(H)$, and $B_1(H)$ for the four and five Tesla fields. We find
$\xi_c=3.78\AA$, $H_{c2}'=1.837 T/K$, $\kappa=56.02$, and $T_c(H)=90.91K$, $90.38K$, $89.83K$,
and $89.04K$ for $H=2T$, $3T$, $4T$, and $5T$ respectively and show the fits in
Fig.~\ref{jeanfig}. The fit to the 5T data is the least satisfactory which we attribute to the
data: One can see from Fig.~3 of Ref.~\onlinecite{jean96} that the 4T and 5T data have spurious
behavior at low temperatures instead of collapsing to the mean-field temperature dependence
which could be a result of the entry into the irreversible region. Except for the 5T fit, the
fits are reasonable and the parameter values are similar to those found by others which gives us
confidence that the theory is a good description of the data. Furthermore, when one compares the
value of $H_{c2}'$ obtained in the fits to the values $H_{c2}'=1.85T/K$ found from the
$T_c(H)$'s for $H=2$-$4T$ (throwing out the less satisfactory $5T$ fit), one finds good
agreement strengthening the credibility of the fit.  If the quantities $Q$, $K$, and $M$ are
added as fitting parameters,  rather than being taken as obtained from the numerical
``calibration'', significantly better fits are not obtained. This is as expected, if the
procedure is correct.

In the Inset to Fig.~\ref{jeanfig}, we show the LLL scaling of the 
magnetization data of Ref.~\onlinecite{jean96} 
using the parameters obtained from our fit. 
The collapse of the data is good in the critical region but fans out 
somewhat
in the low-temperature, mean field region. We 
believe that some of this fanning may be due to the spurious behavior 
associated with the 
irreversible region which we discussed above.\cite{JMdetail} 

We have also attempted fits to the magnetization data
on a YBCO single crystal by Salem-Sugui and da Silva\cite{salem96}.
In this case the fits were rather poor unless unphysical parameter values were
chosen.  The situation did not improve when $Q$, $K$, and $M$ were added as fitting 
parameters. This is in contradistinction with what occurs with the numerical results and with
the data of Ref.~\onlinecite{jean96}. We are more inclined to believe that the origin of
the discrepancy lies with the data set than with the theory, but we cannot be certain until more
magnetization data from YBCO becomes available to us.    

We now turn to using Eq.~(\ref{M}) to explain the results of Ref.~\onlinecite{jean96b} in which
it was found that the temperature derivative of the magnetization has an approximately  field
independent value  and a crossover point at a distinct temperature  whose value lies close to
that of $T_{c0}$ for YBCO. 

\subsection{Behavior of the Magnetization Temperature Derivative}
\label{sec:der}
The behavior of the temperature derivative of the magnetization $\partial M(H,T)/\partial T$ has
been studied as a function of the field $H$ in Ref. \onlinecite{jean96b}. The behavior is quite
striking. The most salient feature of the experimental results (Fig.~2 of Ref.
\onlinecite{jean96b}) is the very weak dependence on the field of this partial derivative at
temperatures near the mean-field temperature. This weak dependence extends to a rather wide
temperature range (more than eight degrees) and in fact, can nearly be called a field
independence for the fields $H\ge 2$ for both the YBCO and BSCCO data. Also remarkable is
the feature upon which the authors in the cited experimental work focused which is the apparent
crossing of the data at a temperature very close to $T_{c0}$.

These experimental results can be easily understood from the LLL scaling formula, Eq.~(\ref{M}).
This equation can be rewritten simply as: \begin{equation} M=(HT)^{2/3}\mu (t), \label{Mu}
\end{equation} where $\mu$ is the scaling function and $t$ the scaled temperature variable
defined in Eq.~(\ref{tsc}). From these equations we have for the temperature derivative:
\begin{equation} \frac {\partial M } {\partial T}= \frac{2}{3} (HT)^{-1/3} \mu (t) +\mu'(t)
(1-\frac{2}{3}\frac{T-T_c(H)}{T}), \label{firstd} \end{equation} where $\mu'(t)$ is the
derivative of $\mu(t)$ with respect to its argument. We can the proceed to the evaluation of the
mixed second partial derivative, with the result: \begin{eqnarray} \frac{\partial^2 M}{\partial
T \partial H}=&&\frac{4}{9}\frac{1}{HT} (\mu(t)(HT)^{2/3}-\mu'(t)(T-T_c(H)))\nonumber\\&& -
\mu"(t)(1-{2\over 3} \frac{T-T_c(H)}{T}) \nonumber\\&&\times
\frac{1}{HT^{2/3}}(\frac{1}{H'_{c2}}+\frac{2}{3}\frac{T-T_c(H)}{H}), \label{mixed}
\end{eqnarray} where we have used $dT_c(H)/dH=1/H'_{c2}.$ Because the scaling function for the
magnetization is essentially a linear function of its argument (except in a very narrow region
near its kink), we can drop the term proportional to $\mu"$ in Eq.~(\ref{mixed}). We then have
that the condition for the mixed derivative to vanish is, \begin{equation}
(HT)^{-1/3}\mu(t)=\mu'(t)\frac{T-T_c(H)}{HT}, \label{cond} \end{equation} which, taking into
account Eq. (\ref{tsc}), can be written \begin{equation} \mu(t)=\mu'(t)t \approx \mu'(0) t.
\end{equation} Since $\mu(0)$ is small, this relation is to good accuracy the Taylor expansion
of $\mu(t)$ about $t=0$ and it will trivially hold over an extended region. Indeed, since $\mu$
is nearly everywhere linear, so that there are no higher order terms, it may appear that we have
nearly proved that the vanishing of the mixed derivative is an identity. This is not at all the
case, chiefly because $\mu(0)$ cannot everywhere be neglected. However, the argument makes it
abundantly clear that the weak dependence on the field of the mixed derivative of $M$ and the
crossing point follow easily from LLL scaling.

\subsection{Specific Heat}
\label{sec:sh}
In this section we will examine the specific heat function [Eq.~(\ref{SH})] 
fitting the data of Ref.~\onlinecite{jean96} to it. We will then use 
Eq.~(\ref{SH}) to address the results of Ref.~\onlinecite{junod96} in which evidence for a
second order flux lattice melting transition is claimed and show how such experimental results
can be alternatively explained via Eq.~(\ref{SH}) without need to invoke flux lattice melting
arguments. 

\begin{figure}[htb]
\centerline{\epsfysize=2.4truein \epsffile{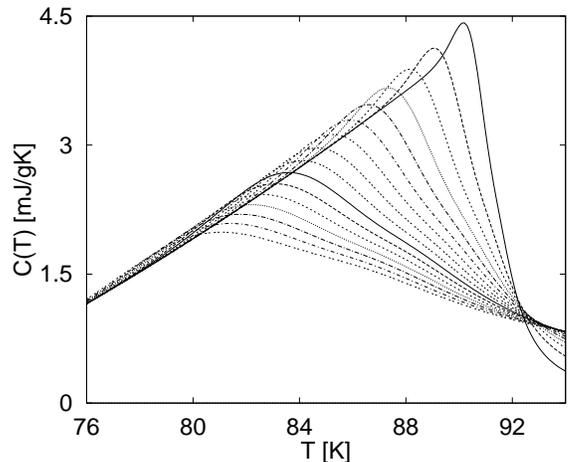}}
\caption[]{\em{ Theoretical specific heat curves for various fields ($H=1$-$8T$
plotted at 0.5 T intervals) calculated from Eq.~(\protect\ref{SH}). }}
\label{shfig}    
\end{figure}

\begin{figure}[htb]
\centerline{\epsfysize=2.4truein \epsffile{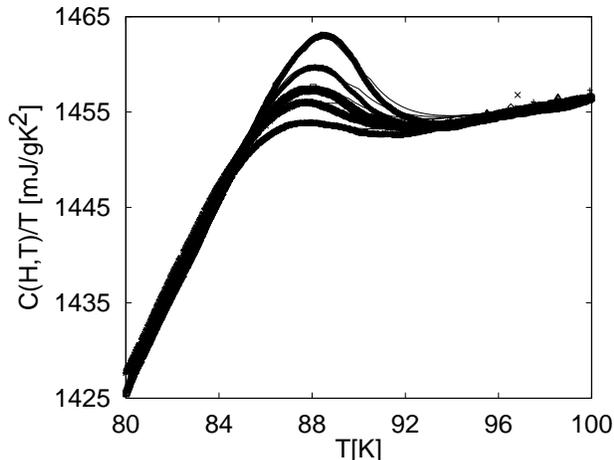}}
\caption[]{\em{ YBCO specific heat data for various fields ($H=3$-$7T$) from
Ref.~\protect\onlinecite{jean96} along with the fits to Eq.~(\protect\ref{SH}).}}
\label{jeanfitfig}    \end{figure}

We begin by examining the specific heat curves produced by Eq.~(\ref{SH}). We display 
this function for various fields in the vicinity of the critical area using parameters 
characteristic of YBCO in Fig.~\ref{shfig}. For $C_{MF}(T)$, we used a
standard form: $C_{MF}(T)=\gamma T[1+b(T/T_{c0}-1)]$ which was taken to be a
constant above $T_c(H)$. This form produces an artificial nonanalyticity of
the curves at $T_c(H)$ since the first derivatives are not continuous but this is unimportant
since it does not affect our analysis in any significant way. As one can see in
Fig.~\ref{shfig}, the curves generated from Eq.~(\ref{SH}), which have the features of a
mean-field ``ramp'' with fluctuations which produce a peak, are qualitatively similar to the
data from YBCO samples. See, for example, Fig.~2 of Ref.~\onlinecite{jean96} (reproduced here 
in Fig.~\ref{jeanfitfig}), Fig.~1 of Ref.~\onlinecite{junod96}, or Fig.~2 of 
Ref.~\onlinecite{zhou93}. What is common to all of these 
curves is that they do not collapse immediately for temperatures below the peak. This is in
agreement with a result derived from LLL theory, as originally pointed out in
Ref.~\onlinecite{sasik95}. There, using the Maxwell relation, $(\partial^2M/\partial
T^2)_H=(\partial C_H/\partial H)_T/T$, it was noted that the left hand side of this equation is
positive for lower temperatures, which means that $C(T,H)$ must increase with increasing
field\cite{sasikdetail}. As we will discuss below (Section \ref{sec:3dxy}), 3DXY theory cannot
account for such behavior. 

We have attempted fits of the specific heat data
of Ref.~\onlinecite{jean96} to Eq.~(\ref{SH}). The best 
test of the theory would be to be able to fit the specific heat data with the parameters
obtained for the magnetization data on the same sample, which was discussed in Section
\ref{sec:mag}. This  could not be done however, and a good fit to the entire relevant
temperature range could not be obtained even when lifting the constraints from the magnetization
fits. We believe this is due to the 3D specific heat function {\it above the peaks} not having
the same qualitative behavior as data from YBCO samples which we believe exhibits 2D behavior
in this region. The experimental side of this statement is
reasonable since 2D fluctuations have been observed in these materials through electronic
transport measurements\cite{2div}. (For evidence in specific heat data, see
Refs.~\onlinecite{pierson96,zhou93}.) The theoretical side of this statement is plausible for
the following reasons. As one can see, the theoretical curves for three dimensions clearly
exhibit a crossover point above the peak. While this crossover can be shifted to higher
temperatures by varying the parameters, it still tends to occur at smaller temperatures than it 
does for the 2D theoretical curves. More  important, the 2D curves decay more quickly to 
zero above the transition. 

To account for the apparent dimensional crossover of the specific heat data, we have done fits
of the 2D function\cite{zlatko94} to the data omitting a  sizeable temperature window
[83.3K:91.5K] around the peak where the 3D behavior is expected to dominate in order to fix the
background ($C_B=(-8.6336+2.2602T)$ mJ/gK). We then do a fit of the 3D function to the 3T, 4T,
5T, 6T, and 7T specific heat data for the temperature range [82K:89K]. The results are good as
can be seen from the parameter values: $H_{c2}'=1.82T/K$, $\kappa=65.6$, and $\xi_c=3.1819$.
This is reinforced by a visual inspection of the fit in Fig.~\ref{jeanfitfig} where the curves
agree quite well in the region we believe to be 3D. One also notices that the curves are on the
high side at temperatures above the peak, which agrees with our earlier statement that the 2D
fluctuations decay more quickly with temperature than do the 3D ones.

The parameter space is large here and it would be impossible to explore all
of it to determine the very best fit. From our investigations however, we are
certain that an improved fit with better parameter values can be
obtained for example by allowing $Q$, $K$, and $M$ to vary. One could also
allow for a quadratic term in the background or even in the mean-field term
which we have extended over a large temperature range. Even without exploring
the large parameter space, we have demonstrated the agreement between the
3D LLL specific heat function [Eq.~(\ref{SH})] and YBCO data.

In the remainder of this subsection, we will discuss more indirect
consequences of LLL theory as applied to the specific heat of YBCO class
materials. We will show how using Eq.~(\ref{SH}) we can give an
alternative explanation for features that have been claimed\cite{junod96} as conclusive
evidence for {\it second order} flux lattice melting without invoking flux 
lattice melting arguments. As we pointed out above, the curves in Fig.~\ref{shfig} produced from
Eq.~(\ref{SH}) reproduce the key features of specific heat data from YBCO samples, such as that
seen in Fig.~1 of Ref.~\onlinecite{junod96}, especially for temperatures below the peak. As the
field increases, the peak in the theoretical curves moves down in temperature and broadens. It
is this general behavior which we claim produces the feature that Roulin {\it et al.} cite as
evidence for {\it second order} flux lattice melting. 

To arrive at their conclusion of 
second-order flux lattice melting those authors\cite{junod96} subtract a data 
set for one field from a data set from a slightly larger field and find a
peak in the ``differential''. This peak
is due to the higher field starting to peak at a slightly lower temperature
due to peak broadening and transition temperature suppression. To see if this peak 
corresponds to a flux lattice melting, they look at the field dependence of the peak temperature
and find that it agrees with the same curve found in Ref.~\onlinecite{welp96} by analysis  of
magnetization and resistivity curves to derive a flux lattice melting line for a YBCO sample.
Although this seems persuasive, we will now show that similar results can be obtained from
Eq.~(\ref{SH}). 

\begin{figure}[htb]
\centerline{\epsfysize=2.4truein \epsffile{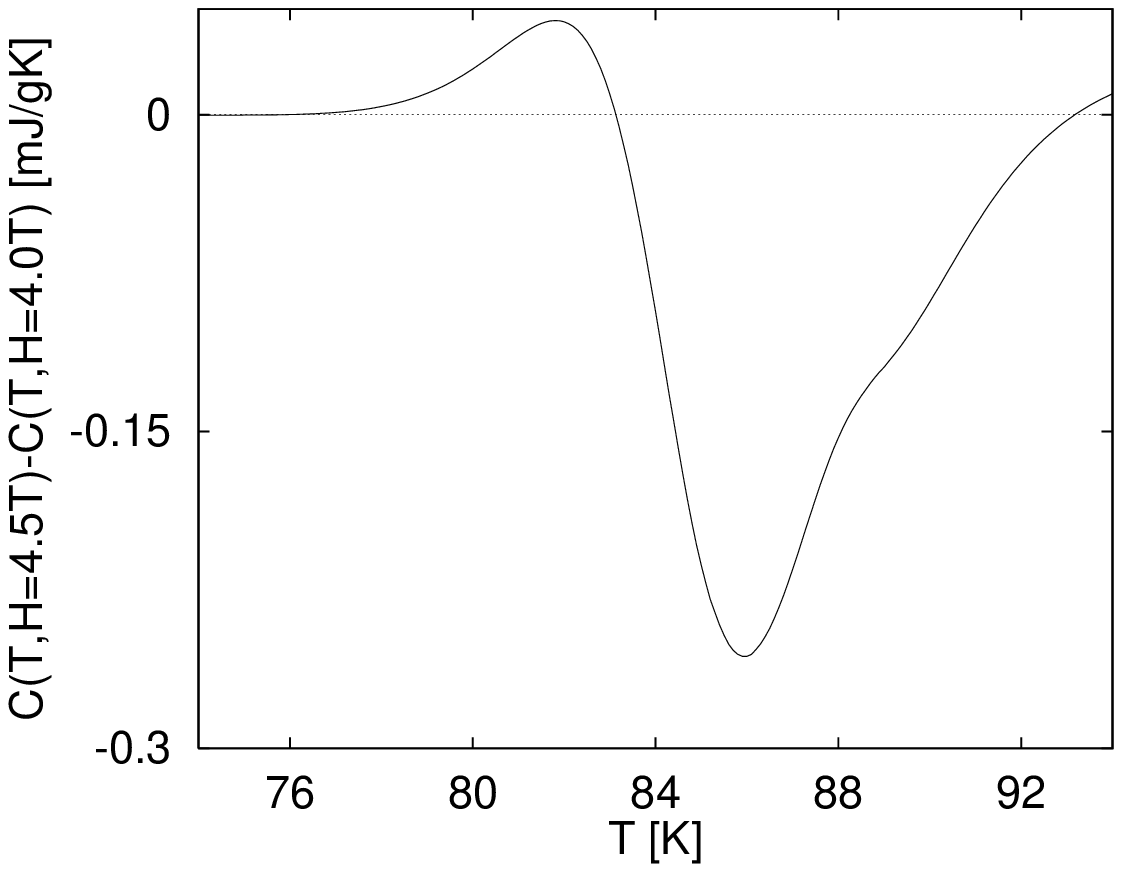}}
\centerline{\epsfysize=2.4truein \epsffile{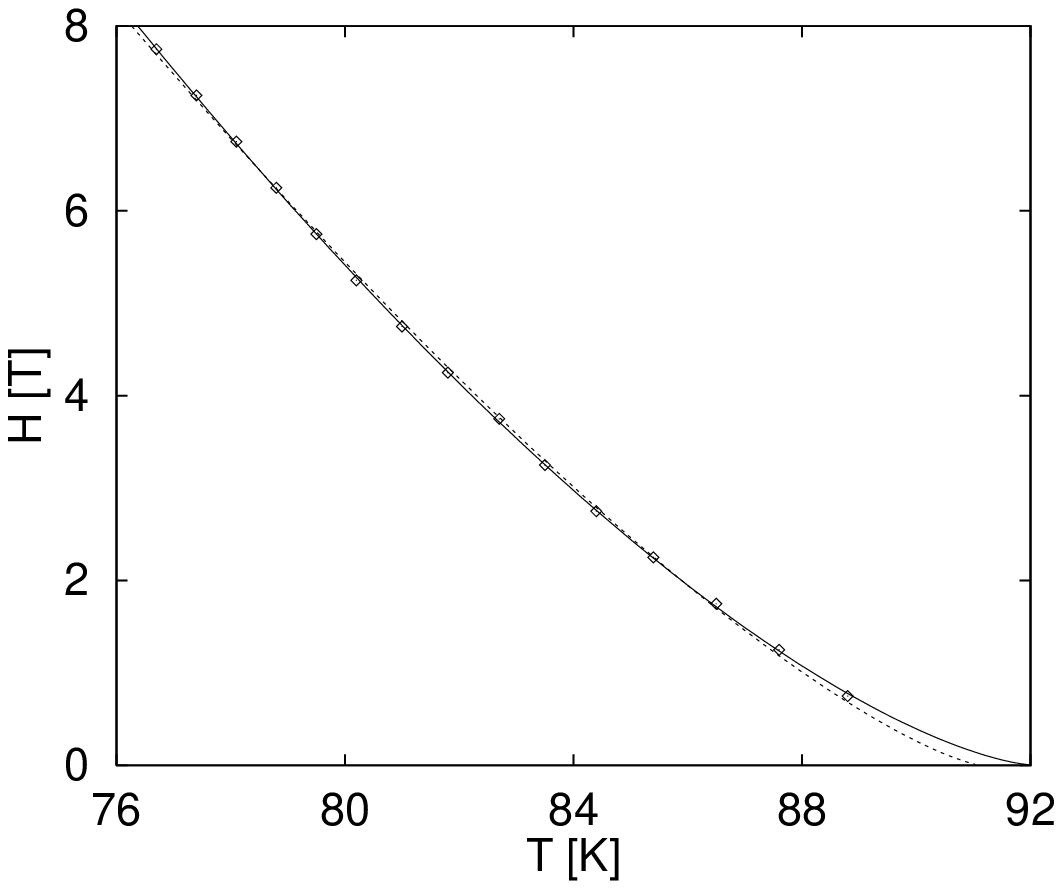}}
\caption[]{\em{(a) A ``differential'' ($\delta C(T,H=4.25T) =$ $C(T,H=4.5T) - 
C(T,H=4.0T)$) from two of the theoretical curves in Fig.~\protect\ref{shfig}. 
This curve has the same qualitative behavior as Fig.~2b of 
Ref.~\protect\onlinecite{junod96}. (b) The field value 
versus the temperature at which the ``differentials'' peak (as explained in the 
text) for those fields and two theoretical fits which are discussed in Section 
\protect\ref{sec:3dxy}.}} 
\label{shdiffig}    \end{figure}

Consider Fig.~\ref{shdiffig}a. Here, $\delta C(T,H=4.25T) = C(T,H=4.5T) - C(T,H=4.0T)$, the 
difference of two theoretical curves calculated from Eq.~(\ref{SH}), is plotted and is 
seen to have the same behavior as that of Fig.~2b in Ref.~\onlinecite{junod96}: a peak 
followed by a deep trough. The 
field dependence of the temperature at which each $\delta C(T,H)$ peaks is then calculated. This
is plotted in Fig.~\ref{shdiffig}b along with two lines. The first line (solid
line) is the best three parameter fit of this theoretical LLL result to the function
$H(T)=a(1-T/b)^c$ with $a=111.2$, $b=92.1$ and $c=1.48$ (standard deviation$=0.02$). The second
(dashed) line is a two parameter fit with a fixed $c=1.33$ and $a=89.2$ and $b=91.1$ (standard
deviation$=0.05$). One can see that the while the exponents differ by 11\%, the two curves are
hard to distinguish. These lines are to be contrasted to that of
Refs.~\onlinecite{junod96,welp96} where the corresponding line, $H_m(T)=99.7[1-T/92.5]^{1.36}$,
is identified with the second order melting line. The numbers in both fits agree quite well with
each other, especially the exponent which is the most critical parameter in this curve. However,
in our case it obviously has nothing to do with melting. The curve
$H_m(T)=99.7[1-T/T_c(0)]^{1.36}$ has not since been experimentally reproduced with the same
numbers. It therefore seems that the agreement between the specific heat data of
Ref.~\onlinecite{junod96} and that of Ref.~\onlinecite{welp96} may be just coincidental, and
reflect a property of the LLL specific heat.

The authors of Ref.~\onlinecite{junod96} have since come out with separate and more 
convincing evidence\cite{junod96b} for a {\it first order} phase transition in a 
cleaner YBCO sample finding data similar to that of Schilling {\it et 
al}\cite{schilling96}. Fig.~2b of Ref.~\onlinecite{junod96b}, the analog of their 
Fig.~2a in Ref.~\onlinecite{junod96}, shows clear spikes that were not present in 
their original paper. Our argument above does not, 
of course, affect this independent evidence for a {\bf first order} flux lattice melting 
transition which now seems to be well documented.\cite{welp96,welp96b}

\subsection{Implications for 3DXY Scaling}
\label{sec:3dxy}
In this subsection, we will address an important question concerning the
experimental discrimination between LLL and 3DXY scaling. We will show that
the LLL theoretical results derived from Eqs.~(\ref{M}) and (\ref{SH}) 
can be used to generate curves which can then apparently be scaled in accordance with 3DXY 
theory. With such a result, one has to question the meaning of 3DXY scaling in 
experimental data: finding 3DXY scaling does not at all exclude that the
data actually is in agreement with GL-LLL theory. 
Further, the importance of comparing the experimental results
with the actual functions is made obvious. This is a problem for 3DXY theory since no scaling 
functions are available for it.

There are two scaling forms that have been used to analyze
specific heat data according to 3DXY scaling theory. The 
first is that derived in Ref.~\onlinecite{salamon93}: 
\begin{equation}
\frac{C(H,T)-C(H=0,T)}{H^{0.0097}}=f(\frac{T/T_c-1}{H^{0.747}})
\label{3dxy1}
\end{equation}
where $T_c$ is the zero-field critical temperature and the second is that derived in 
Ref.~\onlinecite{leeds94a}: 
\begin{equation}
\frac{C_{SC}(H,T)-C_0}{H^{0.0097}}=f(\frac{T/T_c-1}{H^{0.747}})  
\label{3dxy2}
\end{equation}
where $C_0$ is the height of the specific heat cusp and the subscript $SC$ signifies that it is
only the superconducting contribution, with the background subtracted out. (The exponent for $H$
on the right hand side (RHS) of these equations is derived from the specific heat exponent
$\alpha$ and we have used the value derived from $^4$He experiments. The theoretical value of
$\alpha$ is $0.005$.)  The second scaling form\cite{leeds94a} is the more convincing one since
the scaling takes place over a wider  and less trivial range. In the first form on the other
hand, one is scaling mostly horizontal lines which are equal to zero\cite{salamon93}, and thus
are virtually guaranteed to scale. The first form thereby acts as a sufficient condition for
3DXY scaling and we use it here. If it does not scale according to this form, it certainly won't
scale according to Eq.~(\ref{3dxy2}). 

\begin{figure}[htb]
\centerline{\epsfysize=2.4truein \epsffile{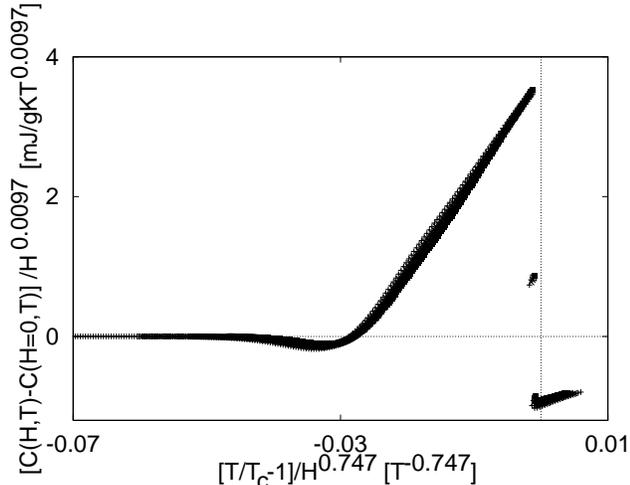}}
\caption[]{\em{ The theoretical specific heat curves from an LLL calculated 
Eq.~(\protect\ref{SH}) (Fig.~\protect\ref{shfig}) scaled according to 3DXY theory: 
Eq.~\protect\ref{3dxy1}.}}
\label{zl3dxyfig}    
\end{figure}

Thus, we proceed to generate ``data'' from the theoretical expression
Eq.~(\ref{SH}). This is a portion of the theoretical results shown
earlier in Fig.~\ref{shfig}. Then, we attempt to scale this ``data''
according to the 3DXY formula.
The scaling results are shown in Fig.~\ref{zl3dxyfig} for the $4T$, $5T$, $6T$, $7T$, and $8T$
fields and as one can see, the collapse is reasonable. We have left out the smaller fields which
as one would expect do not collapse onto these curves. If one were to consider the noise, and
the background subtractions, which inevitably enters enter into the analysis of actual
experimental data, one could call our scaling convincing. (When the exponent $0.005$ is used for
the exponent for $H$ on the RHS of Eq.~(\ref{3dxy1}), the ``data'' does not collapse quite as
well.) That LLL calculated ``data'' can be made to scale according to 3DXY theory exemplifies
the excessive generality of 3DXY scaling and again raises questions  on the significance of it.

\begin{figure}[htb]
\centerline{\epsfysize=2.4truein \epsffile{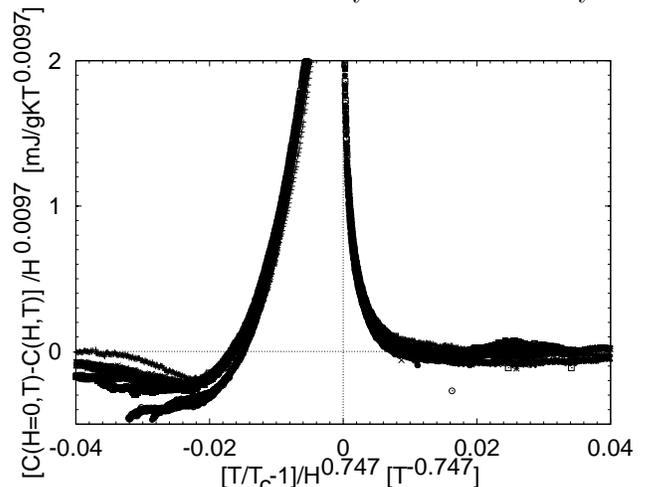}}
\caption[]{\em{ The data of Ref.~\protect\onlinecite{junod96} scaled according to 
3DXY theory. The critical region is to the left of the large peak which is 
the zero-field peak. One can see that 3DXY scaling does not describe the 
critical region. There are seven data sets plotted,
corresponding to fields of 1 to 7 T at 1T intervals. Higher field sets have 
lowers values at the extreme left of the graph.}}
\label{jean3dxyfig}    \end{figure}

As a further insight on the problems with 3DXY scaling, we comment on actual specific heat data
from a YBCO sample which was scaled using Eq.~\ref{3dxy1} the same way as the theoretical curves
in Fig.~\ref{zl3dxyfig}. This is the specific heat data of Ref.~\onlinecite{jean96}, the same
data to which we attempted fits in Section \ref{sec:sh}. The 3DXY scaling results in that work
are shown in Fig.~4 in Ref.~\onlinecite{jean96}. The large central peak is the zero-field peak
and so of course it is unimportant since it is simply an artifact of having subtracted off the
zero-field data. The important region to consider is just to the left of the central peak, since
that area represents the peaks (or the region of critical behavior) of the finite-field data
sets. As one can see, the data is not close to collapsing here. Because this data is plotted on
a scale which brings in the size of the irrelevant, large zero-field peak, scaling does not
appear to be a dramatic failure. However when compared to the collapse of the data at
temperatures above the peak where there are no fluctuations and one is dealing with only the
background, one can see that the failure of 3DXY scaling is more obvious. We have done our own 
similar analysis on this data which we show in Fig.~\ref{jean3dxyfig} in a more restricted
scale. (We emphasize the point that if the data does not scale with Eq.~\ref{3dxy1}, it will not 
scale with the more convincing Eq.~\ref{3dxy2}.) It is seen that the $2T$ and $3T$ data do 
collapse but the $1T$ data rides high and the
larger fields go low. The failure of the scaled data to collapse in the actual critical region
makes one question whether 3DXY behavior is valid. This is, again, one of the reasons why
derivations of the actual functions like those of Ref.~\onlinecite{zlatko92} are so important.

We close this section on implications for 3DXY scaling with a discussion of the 
references to 3DXY in the flux lattice melting literature. There a melting line of 
the form $H_m(T)=99.7[1-T/T_c(0)]^{1.36}$ is found by many 
authors\cite{junod96,welp96,liang96} who point out that the exponent here is the 
nearly the same as that (1.33) expected for 3DXY critical point analysis. And 
while some of the same authors\cite{welp96} note that such an
analysis appears incompatible with a first-order transition, that we get the same
exponent from a LLL approach appears to make such an identification with 3DXY theory 
even more unlikely.

\section{Discussion and Summary}  
\label{sec:summ}

In this paper, we have studied the relevance of 3D GL-LLL
theory as exemplified by Eqs.~(\ref{M}) and (\ref{SH}) 
for the magnetization and specific heat, to the relatively
isotropic HTSC materials of the YBCO family.

Eq.~(\ref{M}) was found to accurately describe the magnetization data
for a YBCO 
sample\cite{jean96} and the numerical data
modeling these materials in the GL-LLL 
formalism\cite{sasik95}.
In the former case, several of the fitting parameters were obtained 
independently from fits to the numerical calculation thereby 
raising the credibility of the fits. The 
remaining parameter values ($H_{c2}'$, $\kappa$, and $\xi_c$) correspond well to those found for
these materials by other means. When applied to another data set, \cite{salem96} 
Eq.~(\ref{M}) was not found to accurately describe it. We obviously can not completely rule
out that the results in the last reference are the correct ones and that the other two are
wrong, but it seems unlikely to us. We also found that Eq.~(\ref{SH}) could not describe the
YBCO specific heat data of Ref.~\onlinecite{jean96} over the whole temperature [80K:100K] but
that very good results for five of the fields could be obtained if the region above the
peak is excluded from the fit. The question of why this is brings us now to a broader discussion
of the applicability of both the 3D specific and magnetization functions of
Ref.~\onlinecite{zlatko94} to YBCO data.

As we discussed in the Introduction, several factors could explain possible
discrepancies between the functions and the data. The most significant would
be the failure of GL-LLL theory to describe the YBCO data. We 
believe that the evidence is against this because it has been shown elsewhere
(as mentioned in the Introduction)
that YBCO data scales according to this theory. Another possible reason would
be discrepancies between the functions of Ref.~\onlinecite{zlatko94} and exact
GL-LLL theory. We tend to discount this for two reasons. First, 
the 2D function was found to have significant success 
in describing the more anisotropic HTSC materials and is known
to be in excellent agreement with numerical simulations
of the 2D GL-LLL theory. Secondly, we note the exceptional 
fit of Eq.~(\ref{M}) to the 
numerical data (Fig.~\ref{sasikfig}a) which provides 
evidence that this equation is an accurate 
description of the GL-LLL theory. Rather, 
we believe that the shortcomings arise because 
Eqs.~(\ref{M}) and (\ref{SH}) are 3D functions and 
it has been shown that, while the YBCO is the 
least anisotropic of the major HTSC materials, 
2D signatures are present\cite{2div}. As mentioned above, 
we have tested this by successfully fitting the 3D 
LLL function to the temperature range of the specific 
heat data which is believed to be 3D. Further 
adding to our conclusion here is prior evidence for 2D behavior in 
YBCO and LBCO through specific heat LLL scaling.\cite{pierson96,zhou93}

The reason why the shortcomings appear to affect the specific heat and not the
magnetization appears to be the following: the specific heat has a more
complicated behavior than the magnetization,
which is monotonic. However, the possibility that the approximations involved
in the computation of Eqs.~(\ref{M}) and (\ref{SH}) are inadequate in the
region above the peak cannot be conclusively ruled out at this point. 
To try to remedy the specific heat problem by 
splicing together the 2D and 3D functions in the appropriate temperature ranges to describe the
YBCO data would introduce so many fitting parameters that any fit would be of little value. It
is also possible to use the quasi-2D functions of Ref.~\onlinecite{zlatko94} but these also have
a large number of fitting parameters besides being much less tractable than the pure 2D or 3D
functions. 

The importance of scaling functions, as contrasted to mere scaling variables, has been 
demonstrated in this paper by showing the low information content of 3DXY scaling without the 
associated scaling functions. It was shown in two cases 
(Section \ref{sec:3dxy}) how curves from an equation 
calculated using LLL assumptions could be 
described by 3DXY theory. This has consequential 
ramifications for the significance of 3DXY 
theory and what it means to find that 
data scales according to 3DXY theory.
In spite of the generality of 3DXY theory, 
it was further demonstrated that it does not describe 
the finite field specific heat data of the YBCO samples in the critical region.

\acknowledgements

Conversations with Dr.~J.~Buan and Prof.~C.~C.~Huang are gratefully acknowledged. We thank
Drs.~\v S\'a\v sik and Stroud, Dr.~Salem-Sugui Jr., {\it et al.}, and Dr.~O.~Jeandupeux {\it et al.} for generously providing their data. We also thank
Isaac Rutel for assistance with the data used in Fig.~\ref{zl3dxyfig} and \ref{jean3dxyfig}.
This work has been supported in part by the NSF Grant No. DMR-9415549.


\begin{references}
\bibitem{zldetail} It should be noted that in the equation for the 3D specific heat function
(Eq. (27)) in Ref.~\protect\onlinecite{zlatko94}, only leading terms were included. In general,
the subleading terms are needed to reproduce the peak in the 3D LLL specific heat function.
Such terms can be extracted from the 3D limit of the general expression (Eq. (30)) in
Ref.~\protect\onlinecite{zlatko94}.
\bibitem{q3d} The resulting expression is more accurate than
that quoted for the pure 3D limit in Ref.~\protect\onlinecite{zlatko94} which includes leading
terms only.
\bibitem{back} The data were made available to us with a background term
already subtracted in connection with a different fitting, the subtraction discussed here refers
therefore to an adjustment in the background term.
\bibitem{JMdetail} It should be noted that our
3D LLL scaling of this data is an improvement of that done on this data by the authors of
Ref.~\protect\onlinecite{jean96} which is shown in their Inset to Fig.~5 over a more restricted
scale.
\bibitem{sasikdetail} At temperatures just below the transition temperature, it was observed by
the authors of Ref.~\protect\onlinecite{sasik95} that $(\partial^2M/\partial T^2)_H$ is negative
implying that $(\partial C_H/\partial H)_T$ must also be negative as observed in
Fig.~\ref{shfig}. It is further noted that this region corresponds to the vortex liquid phase
and that when $(\partial^2M/\partial T^2)_H$ is positive, the region is a vortex solid phase.
This appears to be a coincidence in so for as that no connection can be made between the
temperature at which $(\partial^2M/\partial T^2)_H=0$ and a melting temperature.
\end{references}
\end{document}